\documentclass[aps,preprint,preprintnumbers,amsmath,amssymb]{revtex4}
\usepackage{amssymb}
\usepackage{graphicx}
\usepackage{dcolumn}
\usepackage{bm}

\begin{document}

\title{Antiferromagnetic Kondo lattice compound CePt$_{3}$P}

\author{Jian Chen$^{1,2,}$\footnote[1]{Electronic address: chenjian123@zju.edu.cn}, Zhen Wang$^{1}$, Shiyi Zheng$^{1}$, Chunmu Feng$^{1}$, Jianhui Dai$^{3}$, and Zhuan Xu$^{1,4,5,}$\footnote[2]{Electronic address: zhuan@zju.edu.cn}}

\affiliation{
$^1$State Key Lab of Silicon Materials and Department of Physics, Zhejiang University, Hangzhou 310027, China\\
$^2$Zhejiang University of Water Resources and Electric Power, Hangzhou 310018, China\\
$^3$Department of Physics, Hangzhou Normal University, Hangzhou 310036, China\\
$^4$Zhejiang California International NanoSystems Institute, Zhejiang University, Hangzhou 310027, P. R. China\\
$^5$Collaborative Innovation Centre of Advanced Microstructures, Nanjing 210093, P. R. China
}

\date{\today}
\maketitle

\textbf{A new ternary platinum phosphide CePt$_{3}$P was synthesized and
characterized by means of magnetic, thermodynamic and transport
measurements. The compound crystallizes in an antiperovskite
tetragonal structure similar to that in the canonical family of
platinum-based superconductors $A$Pt$_{3}$P ($A$ = Sr, Ca, La) and
closely related to the noncentrosymmetric heavy fermion
superconductor CePt$_{3}$Si. In contrast to all the superconducting
counterparts, however, no superconductivity is observed in
CePt$_{3}$P down to 0.5 K. Instead, CePt$_{3}$P displays a
coexistence of antiferromagnetic ordering, Kondo effect and
crystalline electric field effect. A field-induced spin-flop
transition is observed below the magnetic ordering temperature
$T_{N1}$ of 3.0 K while the Kondo temperature is of similar
magnitude as $T_{N1}$. The obtained Sommerfeld coefficient of
electronic specific heat is $\gamma_{Ce}$ = 86
mJ/mol$\cdot$K$^{2}$ indicating that CePt$_{3}$P is a moderately
correlated antiferromagnetic Kondo lattice compound.
}\\

\flushbottom
\maketitle
%
%


\section*{Introduction}

The interplay among spin, charge and orbital degrees of freedom in
transition metal compounds has triggered enormous research
interests in condensed matter physics and material science. For a
large family of layered $3d$ electron superconductors (SCs) such
as the copper oxides \cite{Cu} and iron pnictides \cite{Fe}, the
spin fluctuations caused by strong $3d$ electron correlations play
a vital role in the unconventional superconductivity. Besides
these $3d$ transition metal systems, several platinum-based SCs
exhibit remarkably rich physical properties and therefore have
also attracted considerable attention, partly owing to the
moderately strong spin-orbit coupling of  the platinum $5d$
electrons. The most prominent example is the heavy fermion
noncentrosymmetric (NCS) SC CePt$_{3}$Si, in which exotic
superconductivity is observed below $T_{c}$ = 0.75 K
\cite{CePt3Si}: an admixture of spin-singlet and spin-triplet
pairing symmetry, nodal gap structure and huge upper critical
field ($B_{c2}$ $\approx$ 4 T) \cite{NCSbook}. The delicate
interplay between the cerium $4f$ and the platinum $5d$ electrons
places this material on the border of the magnetic quantum
critical point (QCP) but still in the antiferromagnetic (AFM)
gound state, rendering the role of inversion symmetry unclear
\cite{Frigeri}. Among a series of filled skutterudite
$MT_{4}X_{12}$ ($M$ = rare-earth or alkaline-earth metals, $T$ =
transition metals and $X$ = P, As, Sb and Ge) with the cubic space
group $Im\bar{3}$ (No.204), PrPt$_{4}$Ge$_{12}$ was reported to
exhibit time-reversal symmetry breaking from zero-field $\mu$SR
measurements \cite{PrPt4Ge12}. As a result of the unexpectedly
high transition temperature $T_{c}$ = 7.9 K and the moderately
enhanced Sommerfeld coefficient $\gamma$ = 76
mJ/mol$\cdot$K$^{2}$, PrPt$_{4}$Ge$_{12}$ has been extensively
studied and multiband superconductivity has been proposed based on
the analysis of the photoemission spectroscopy \cite{Nakamura} as
well as the magnetic penetration depth \cite{Zhang}. Moreover,
SrPtAs is recently reported to crystallize in a hexagonal
structure ($P6_{3}/mmc$, No.194) with weakly coupled PtAs layers
forming a honeycomb lattice \cite{Nishikubo}. The peculiar locally
NCS structure within PtAs layer together with a strong spin-orbit
coupling demonstrates SrPtAs as an attractive material to explore
superconductivity with a spontaneous static magnetic field
$B_{s}$ \cite{Biswas}.

It is interesting that among the platinum-based superconductors,
the newly reported family of $A$Pt$_{3}$P ($A$ = Ca, Sr and La)
shares the structural similarity with that of iron pnictides
\cite{Takayama}. These compounds crystallize in a tetragonal
structure with space group $P4/nmm$ (No.129) with stacking in the
order of $A$-Pt$_{6}$P-$A$ along the $c$-axis. The distorted
antiperovskite Pt$_{6}$P octahedral unit alternates within the
$ab$ plane, forming an antipolar pattern. The $z$ $\rightarrow$
$-z$ inversion operation is thus preserved. Due to the structural
distortion, the platinum atoms take two different sites as Pt(I)
and Pt(II) so that the Pt(II) and P atoms form a Pt$_{2}$P layer
resembling the FeAs layer in the iron-based superconductors. Of
course, the structure of $A$Pt$_{3}$P is also somewhat similar to
that of CePt$_{3}$Si, but the latter is actually isotypic to the
NCS compound CePt$_{3}$B with the space group $P4mm$ (No.99)
\cite{CePt3Si}. The corresponding Pt$_{6}$Si unit has the polar
structure  under this space group leading to the absence of
inversion symmetry, different from the antipolar structure in
$A$Pt$_{3}$P. Noticeably, the $A$Pt$_{3}$P family shows a
significant variation of $T_{c}$, i.e., $T_{c}$ = 8.4 K, 6.6 K
and 1.5 K for $A$ = Sr, Ca and La, respectively. It was reported
theoretically that spin-orbit coupling (SOC) effect is significant
in LaPt$_{3}$P but negligible in CaPt$_{3}$P and SrPt$_{3}$P
\cite{Chen, Kang, Subedi}. The origin of significantly enhanced
$T_{c}$ in SrPt$_{3}$P is still debatable. It was suggested to be
due to a possible dynamic charge-density-wave (CDW) \cite{Chen}.
However, a theoretical work by Zocco et al. indicated SOC could
strongly renormalize the electron-phonon coupling of SrPt$_{3}$P
and thus enhance the electronic density of states near the Fermi
level \cite{Zocco}. Moreover, several theoretical works claimed that
the CDW instability could not be reproduced in SrPt$_{3}$P \cite{Kang,
Subedi}. The centrosymmetric (CS) compounds $A$Pt$_{3}$P reported
so far do not involve the $4f$ electrons. The interplay between
strong $4f$ electron correlation and superconductivity of $5d$
electrons in the $A$Pt$_{3}$P family remains an open issue.

In this paper, we report our successful synthesis of such a
candidate compound CePt$_{3}$P in the platinum-based phosphides
$A$Pt$_{3}$P family. We performed systematic measurements of the
physical properties including the magnetic susceptibility,
magnetization, specific heat and electrical resistivity. However,
no evidence of superconductivity is observed down to 0.5 K in
CePt$_{3}$P, in contrast to other $A$Pt$_{3}$P compounds. Instead,
the compound displays the rich physics involving the coexistence
of magnetic ordering, Kondo coherence as well as crystalline
electric field (CEF) effect. We shall discuss these properties and
highlight the delicate $4f$-$5d$ interplay in this system.

\section*{Results and discussion}

\begin{figure}[ht]
\centering
\includegraphics[width=10cm]{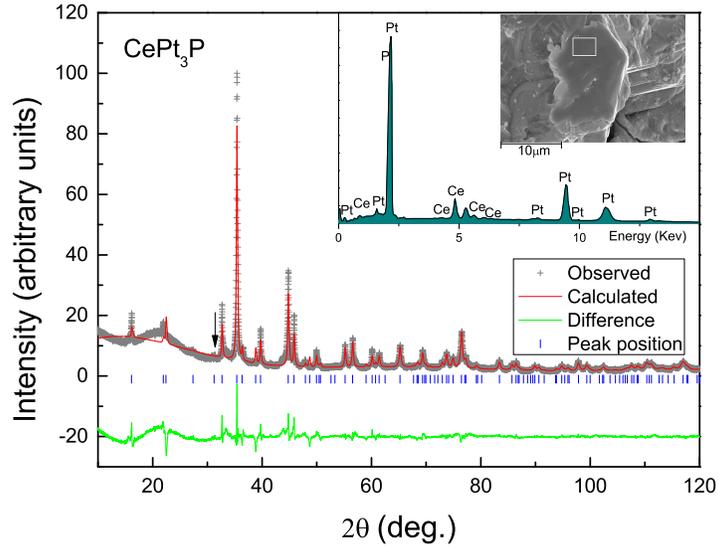}
\caption{\textbf{Rietveld refinement of the polycrystalline CePt$_{3}$P XRD pattern.} Arrow marks an impurity phase which might be PtP$_{2}$. Inset shows a typical energy-dispersive x-ray spectrum with electron beams focused on the selected area of the as-grown sample.}
\label{fig1}
\end{figure}

Figure \ref{fig1} shows the Rietveld refinement of the XRD pattern
of polycrystalline CePt$_{3}$P samples. Almost all peaks can be well
indexed with the tetragonal structure with the space group $P4/nmm$
(No.129), except for a tiny peak of of an impurity phase around
31.4$^{\rm{o}}$ which might be PtP$_{2}$. The result of the Rietveld
refinement \cite{RIETAN} shows a good convergence: $R_{wp}$ = 13.4\%,
$S$ = 3.3. The refined lattice arameters of CePt$_{3}$P are $a$ =
5.7123(7) ${\rm{\AA}}$ and
$c$ = 5.4679(6) ${\rm{\AA}}$ as listed in Table \ref{tab}. The room
temperature XRD patterns of LaPt$_{3}$P are also refined with $R_{wp}$
= 14.9\%, $S$ = 2.7 (data not shown). The refined lattice parameters
of LaPt$_{3}$P are $a$ = 5.7597(3) ${\rm{\AA}}$ and $c$ = 5.4736(3)
${\rm{\AA}}$. For comparison, the lattice parameters of the other
$A$Pt$_{3}$P compounds are also provided in Table \ref{arttype}.
Onecan see obviously that $a$ of CePt$_{3}$P is smaller, while $c$ is
larger, compared with the lattice parameters of SrPt$_{3}$P. Due to
the lanthanide contraction, both of $a$ and $c$ of CePt$_{3}$P are
smaller than those of LaPt$_{3}$P. From the EDS measurements, the
molar ratio is Ce:Pt:P = (1.0$\pm$0.1):(3.2$\pm$0.2):(0.7$\pm$0.2)
for CePt$_{3}$P and La:Pt:P = (1.0$\pm$0.1):(2.6$\pm$0.1):
(0.8$\pm$0.1) for LaPt$_{3}$P. The actual chemical compositions are
close to the nominal ones, while there seems a deficiency on the P
site for both CePt$_{3}$P and LaPt$_{3}$P.

\begin{figure}[ht]
\centering
\includegraphics[width=9cm]{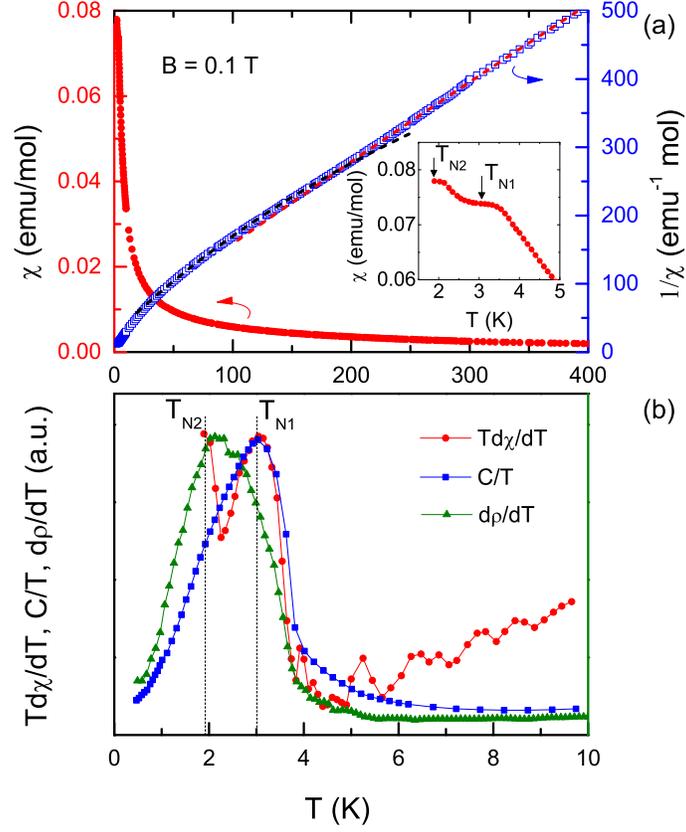}
\caption{\textbf{(a)} Temperature dependence of magnetic
susceptibility, $\chi$, and inverse magnetic susceptibility,
$1/\chi$, of CePt$_{3}$P measured under magnetic field $B$ = 0.1 T
on the left and right axis, respectively. Two dashed lines show
the Curie-Weiss fit for $T$ $>$ 200 K and $T$ $<$ 100 K,
respectively. Inset: enlarged plot of $\chi$ at $T$ $<$ 5 K.
\textbf{(b)} The AFM transition temperature $T_{N1}$ and $T_{N2}$ determined
from the derivative susceptibility $T$d$\chi$/d$T$, specific heat
$C(T)/T$ and derivative resistivity d$\rho$/d$T$ (The complete
data of specific heat and resistivity will be shown in the
following figures).}
\label{fig2}
\end{figure}

The temperature-dependent molar magnetic susceptibility $\chi(T)$
= $M/H$ and inverse magnetic susceptibility 1/$\chi(T)$ of
CePt$_{3}$P measured at $H$ = 1000 Oe are presented in Figure
\ref{fig2}(a). $\chi(T)$ obeys a modified Curie-Weiss law above
200 K, $\chi$ = $\chi_{0}$ + ${\cal C}/(T-\theta)$. $\chi_{0}$ is
a temperature independent susceptibility from the core
diamagnetism, the van Vleck and Pauli paramagnetism, ${\cal C}$ is
the Curie constant and $\theta$ is the Weiss temperature. The
relatively large absolute value of $\theta$ = -28.3 K may be
attributed to the hybridization of the 4$f$ electronic states with
the conduction band \cite{theta_CW}. The derived effective moment
$\mu_{eff}$ = 2.52$\mu_{B}$ is almost equal to that of a free
Ce$^{3+}$ ion, indicating the trivalent Ce ion and well localized
moment of Ce-4$f^1$ electrons at high temperature. $\chi_{0}$ is
in the magnitude order of 10$^{-3}$. For $T$ $<$ 100 K, a change
of the slope of 1/$\chi(T)$ can be clearly observed and the
fitting parameters are $\mu_{eff}$ = 2.11$\mu_{B}$, and $\theta$ =
-15.3 K. Here the change of the slope and the decreased value of
$\mu_{eff}$ can be ascribed to the CEF effect. With decreasing
temperature, $\chi(T)$ increases and shows a round peak around 3.0
K. Upon further cooling, another anomaly is observed near our
base temperature. Two magnetic transition temperatures are
determined from the peaks of derivative susceptibility
$T$d$\chi$/d$T$ as $T_{N1}$ = 3.0 K and $T_{N2}$ = 1.9 K (seen
from Figure \ref{fig2}(b)). Considering the negative Weiss
temperature, the first anomaly marks the AFM ordering below
$T_{N1}$ which is compatible with the magnetization measurement
(discussed below). While the second anomaly is attributed to
a spin-reorientation. A similar phenomenon was observed in
CeNiAsO \cite{CeNiAsO}. Further experimental studies, especially
neutron diffraction measurement on single crystals of CePt$_{3}$P,
are necessary to clarify the magnetic structure at low
temperature.

\begin{figure}[ht]
\centering
\includegraphics[width=10cm]{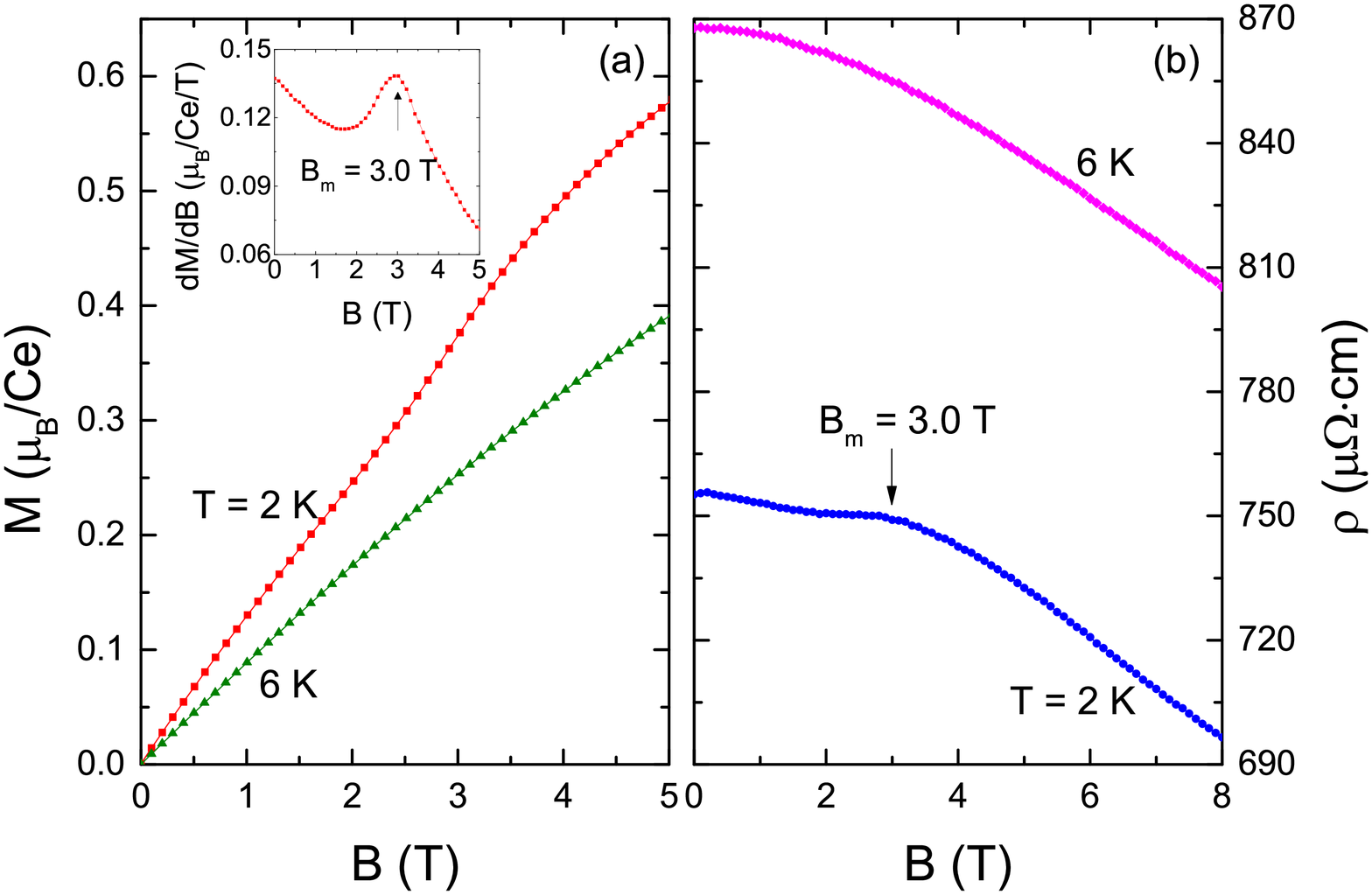}
\caption{\textbf{(a)} Field dependence of the magnetization
$M(B)$. \textbf{(b)} Resistivity $\rho$ of CePt$_{3}$P vs. $B$ measured at
$T$ = 2 and 6 K. Inset to (a) displays the derivative of the
magnetization with respect to the field $dM/dB$ for $T$ = 2 K. \label{fig3}}
\end{figure}

The isothermal magnetization $M(B)$ of CePt$_{3}$P, measured in
the $B$-sweep mode containing both field-up and down loops, is
displayed in Figure \ref{fig3}(a). In the AFM ordering state,
$M(B)$ displays a linear field dependence when $B$ $<$ 2.0 T, but
undergoes a weak step-like increase around 3.0 T. This anomaly,
which is ascribed to a field-induced metamagnetic transition
(MMT), can be independently determined to be $B_{m}$ = 3.0 T by
the peak in $dM/dB$ curve (inset to Figure \ref{fig3}(a)) and the
hump in $\rho(B)$ curve (Figure \ref{fig3}(b)) measured at $T$ = 2
K. The expected hysteresis around $B_{m}$ is not observed and such
absence of hysteresis around MMT was also reported in the
single-crystalline samples CeAuSb$_2$ \cite{MMT1} and YbNiSi$_3$
\cite{MMT2}. No hysteresis in resistivity is observed for
CePt$_{3}$P in this magnetic field range either. Note that the
$M(B)$ curve does not show a saturation trend in the highest field
limit and the value $M$ $\sim$ 0.6$\mu_{B}$ at $B$ = 5 T is much
lower than the theoretical value of 2.14$\mu_B$ for the saturated
moment of free Ce$^{3+}$ ions which is probably due to the CEF
effect. Figure \ref{fig3}(b) shows the isothermal resistivity
versus the applied field. $\rho$ decreases monotonously with
increasing magnetic field at $T$ = 6 K $>$ $T_{N1}$. Whereas at
$T$ = 2 K $<$ $T_{N1}$, a hump around $B_{m}$ = 3.0 T is added to
the decreasing trend. This feature is compatible with the MMT
observed in the magnetization measurement.

\begin{figure}[ht]
\centering
\includegraphics[width=9cm]{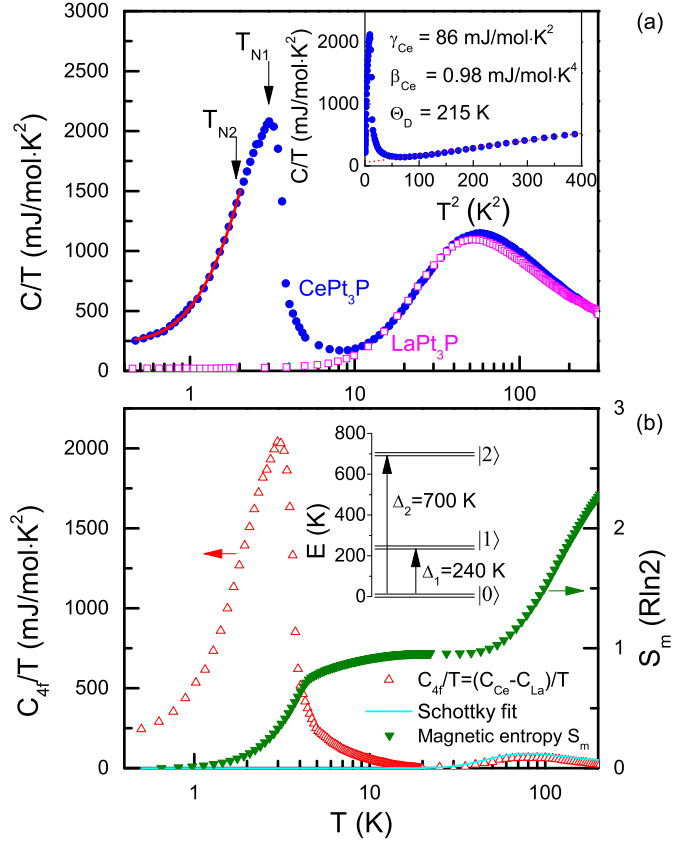}
\caption{\textbf{(a)} Specific heat divided by temperature,
$C/T$, versus log$T$. The solid symbols are for CePt$_{3}$P, while
the open symbols represent the non-magnetic compound LaPt$_{3}$P.
The solid line is a fit to Eq.\ref{eq5} for $T$ $\leq$ 1.9 K.
\textbf{(b)}
The Ce-4$f$ contribution, $C_{4f}/T$, and the magnetic entropy, $S_m$,
on the left and right axis, respectively, measured at zero magnetic
field plotted in a logarithmic temperature scale for $T$ = 0.4-200
K. The solid line shows the Schottky anomaly contribution $C_{Sch}$.
Inset to (a) shows $C/T$ versus $T^2$ together with the fitting
parameters for CePt$_{3}$P (see the text): the Sommerfeld
coefficient $\gamma_{Ce}$, $\beta_{Ce}$ and the Debye temperature
$\Theta_{D}$. The dashed line is a linear fit in the temperature
range $T$ = 10-20 K. Inset to (b) displays the schematic sketch of
CEF energy levels for Ce$^{3+}$ ion in CePt$_{3}$P. \label{fig4}}
\end{figure}

The specific heats of CePt$_{3}$P and LaPt$_{3}$P divided by $T$,
$C(T)/T$, are plotted in the main panel of Figure \ref{fig4}(a) in a
semi-logarithm scale. At room temperature, $C(T)$ saturates to
about 135 and 140 J/mol$\cdot$K for La and Ce compound,
respectively, which are, within an acceptable error range,
compatible with the classical Dulong-Petit law 3$NR$ with $N$ = 5
and $R$ = 8.31 J/mol$\cdot$K, where $R$ is the universal gas
constant. The specific heat $C(T)$ of LaPt$_{3}$P is typical for
nonmagnetic metals since no typical anomaly can be observed at
high temperature. At low temperature, the specific heat of
LaPt$_{3}$P is dominated by the electronic and phonon
contributions for $T$ $<$ $\Theta_{D}/10$, therefore, it can be
fitted to a power law $C/T$ = $\gamma_{La}$ + $\beta_{La}T^2$ over
10-20 K (data not shown). Here $\Theta_{D}$ is the Debye
temperature, and $\gamma_{La}$ and $\beta_{La}$ denote the
coefficients of the electronic and phonon contributions,
respectively. It should be noted that there is a small jump around
1 K in the specific heat of LaPt$_{3}$P which should correspond to
a superconducting transition though it is too small to observe in
Figure \ref{fig4}.

In the paramagnetic region above the magnetic transition, the
specific heat of CePt$_{3}$P can be expressed as
\begin{equation}
C = \gamma_{Ce}T + \beta_{Ce}T^3 + C_{Sch}, \label{eq1}
\end{equation}
where the coefficients $\gamma_{Ce}$ and $\beta_{Ce}$ are of
electronic and phonon contributions of CePt$_{3}$P, respectively,
while $C_{Sch}$ describes the Schottky anomaly item. A linear
$T^2$-dependence is clearly seen in $C/T$ vs $T^2$ plot for
temperature below 20 K (see inset to Figure \ref{fig4}(a)).
The derived Sommerfeld coefficient is $\gamma_{Ce}$ =
(86 $\pm$ 1) mJ/mol$\cdot$K$^2$. The value is moderately
enhanced by a factor of 57 compared with that of LaPt$_{3}$P where
$\gamma_{La}$ = (1.5 $\pm$ 0.1) mJ/mol$\cdot$K$^2$, manifesting the
correlation effect contributed from the Ce-4$f$ electrons.
Therefore, CePt$_{3}$P is a Kondo lattice compound due to
the strong $4f$ electron correlation and moderate effective
$4f-5d$ hybridization. Note that $\gamma_{La}$ for LaPt$_{3}$P
derived here is slightly smaller but still in the same magnitude
order with that obtained in Ref.\cite{Takayama}. The reported
phonon coefficients are in reasonable agreement with each other:
$\beta_{Ce}$ = 0.98(1) mJ/mol$\cdot$K$^4$ for CePt$_{3}$P and
$\beta_{La}$ = 0.94(1) mJ/mol$\cdot$K$^4$ for LaPt$_{3}$P, indicating
similar phonon contributions. The Debye temperature $\Theta_{D}$
estimated by using $\Theta_{D}=(12\pi^4NR/5\beta)^{1/3}$ is
(215 $\pm$ 1) K for CePt$_{3}$P and (218 $\pm$ 1) K for LaPt$_{3}$P,
implying that the above analysis is quite self-consistent.

The Ce-4$f$ contribution to the specific heat of CePt$_{3}$P is
then deduced by subtracting the measured specific heat of the
nonmagnetic isostructural reference sample LaPt$_{3}$P from the
total specific heat of CePt$_{3}$P, i.e., $C_{4f}$ = $C_{Ce}$ -
$C_{La}$. The result is shown in the main panel of Figure
\ref{fig4}(b), plotted as $C_{4f}/T$ vs $T$ in a logarithmic
scale. The Schottky anomaly, which is visible as a broad peak
centered around 90 K in $C_{4f}/T$ curve, should be caused by the
excitations between different CEF levels. The Schottky anomaly
with three Kramers doublets (one doublet ground state and two
excited doublets) for Ce$^{3+}$ ion with $j$ = 5/2 experiencing a
tetragonal crystal-field potential can be expressed by
\cite{CEF1,CEF2}
\begin{align}
&C_{Sch}=\frac{R}{g_0+g_1\exp(-\Delta_1/T)+g_2\exp(-\Delta_2/T)}\nonumber\\
&\times\{g_0g_1(\Delta_1/T)^2\exp(-\Delta_1/T)+g_0g_2(\Delta_2/T)^2\exp(-\Delta_2/T)+g_1g_2[(\Delta_2-\Delta_1)/T]^2\exp[-(\Delta_1+\Delta_2)/T]\}.
\label{eq2}
\end{align}
Here $g_i$ = 2 is the degeneracy of the $i$th doublet state and
$\Delta_i$ is the energy difference between the ground state and
the $i$-th excited state (see the schematic sketch drawn in the
inset of Figure \ref{fig4}(b)). Eq.\ref{eq2} is applied
to $C_{4f}/T$ of CePt$_{3}$P over a temperature range of 50-130 K.
The derived CEF energy differences are $\Delta_1$ = (20.9 $\pm$
0.1) meV ($\sim$ (240 $\pm$ 1) K) and $\Delta_2$ = (60.9 $\pm$
0.3) meV ($\sim$ (700 $\pm$ 3) K). This result may explain the
slope change in 1/$\chi(T)$ curve as well as the broad hump in
both $\rho_{mag}$ and $S$. Furthermore, the large value of
$\Delta_{1}$ is consistent with the reduced effective Ce moment
below 100 K. The magnetic entropy gain $S_m$ is calculated by
integrating $C_{4f}/T$ over $T$ and plotted on the right axis in
Figure \ref{fig4}(b). One can see that $S_m$ reaches about
0.51$R$ln2 at $T_{N1}$ and $R$ln2 is recovered at $\sim$50 K,
indicating that the ground state with the AFM ordering of
Ce$^{3+}$ moments is Kramers two-fold degenerate. The plateau over
the temperature range of $T$ = 10-30 K indicates that the first
excited CEF level is far above $T_{N1}$. $S_m$ reaches $R$ln4 at
$\sim$150 K and increases substantially above the Schottky
anomaly. For a Kondo lattice, the Kondo temperature can be
estimated by the magnetic entropy at $T_N$ via \cite{Yashima}
\begin{equation}
S_m(\xi) = R{\ln[1+\exp(-\xi)]+\xi\frac{\exp(-\xi)}{1+\exp(-\xi)}}, \label{eq3}
\end{equation}
where $\xi$ = $T_K/T_N$. The yielded $T_K$ is about (6.1 $\pm$ 0.1) K for CePt$_{3}$P.

At low temperature, $C_{4f}/T$ shows a pronounced $\lambda$-shape
peak at $T_{N1}$ = 3.0 K, implying a second-order phase transition.
The expected jump in specific heat is $\delta C_{4f|T=T_{N1}}$
$\sim$ 6 J/mol$\cdot$K. A slight slope change in $C_{4f}/T$ is
also observed around $T_{N2}$ = 1.9 K, consistent with the
low-temperature anomaly observed in aforepresented $\chi(T)$ curve.
Based on the mean-field theory of Besnus \textit{et al}.
\cite{Besnus} and Bredl \textit{et al}. \cite{Bredl}, the specific
heat jump $\delta C_{|T=T_{N}}$ is related to the Kondo temperature
$T_{K}$ by the following formula
\begin{equation}
\delta C(\zeta) = \frac{6R}{\psi'''(\frac{1}{2}+\zeta)}
[\psi'(\frac{1}{2}+\zeta)+\zeta\psi''(\frac{1}{2}+\zeta)]^2.\label{eq4}
\end{equation}
Here $\zeta$ = ($T_K/T_N$)/2$\pi$, $\psi$ denotes the digamma
function and $\psi'$, $\psi''$ and $\psi'''$ are the first three
derivatives of $\psi$. Then the Kondo temperature can be also
estimated by applying Eq.\ref{eq4}, obtaining a ratio of $T_{K}/T_{N1}$
= 0.88, or $T_{K}$ $\sim$ (2.7 $\pm$ 0.1) K. Therefore, based on both magnetic
entropy and specific heat jump, it is reasonable to estimate $T_K$
$\sim$ 2-6 K in this compound.

In the magnetically ordered state, the AFM spin-wave spectrum
follows a dispersion relation of $\epsilon_{k}$ =
$\sqrt{\Delta^2+Dk^2}$. Here $\epsilon_{k}$ is the excitation
energy, $\Delta$ is the gap in the spin-wave spectrum, and $D$
is the spin-wave stiffness. The phonon contribution,
$\beta_{Ce}T^3$ item, can be subtracted from the total specific
heat $C$ as $\Delta C$ = $C$ - $\beta_{Ce}T^3$. At low
temperature, $\Delta C$ is described by the following expression
\cite{SDW1,SDW2}:
\begin{align}
\Delta C(T)=\gamma_0T+A_{C}\Delta^4\sqrt{\frac{T}{\Delta}}
[1+\frac{39}{20}(\frac{T}{\Delta})+\frac{51}{32}(\frac{T}{\Delta})^2]\exp(-\Delta/T),
\label{eq5}
\end{align}
where the coefficient $A_{C}$ is proportional to $D^{-3}$. Fitting
the specific heat below $T_{N2}$ (solid line in Figure \ref{fig4}(a)) gives
the fitting parameters  $\gamma_0$ = 247 mJ/mol$\cdot$K$^2$,
$\Delta$ = 2.6 K, and $A_{C}$ = 67.5 mJ/mol$\cdot$K$^4$. The
considerably enhanced zero-temperature Sommerfeld coefficient
$\gamma_0$ is about 3 times of $\gamma_{Ce}$ obtained in the
paramagnetic state, indicating the formation of moderate-heavy
quasiparticles in the antiferromagnetically ordered state. It
is worthwile noting that the obtained spin-wave gap $\Delta$
is of the order of magnitude often found in cerium intermetallics
with AFM ground states \cite{theta_CW}.

\begin{figure}[ht]
\centering
\includegraphics[width=9cm]{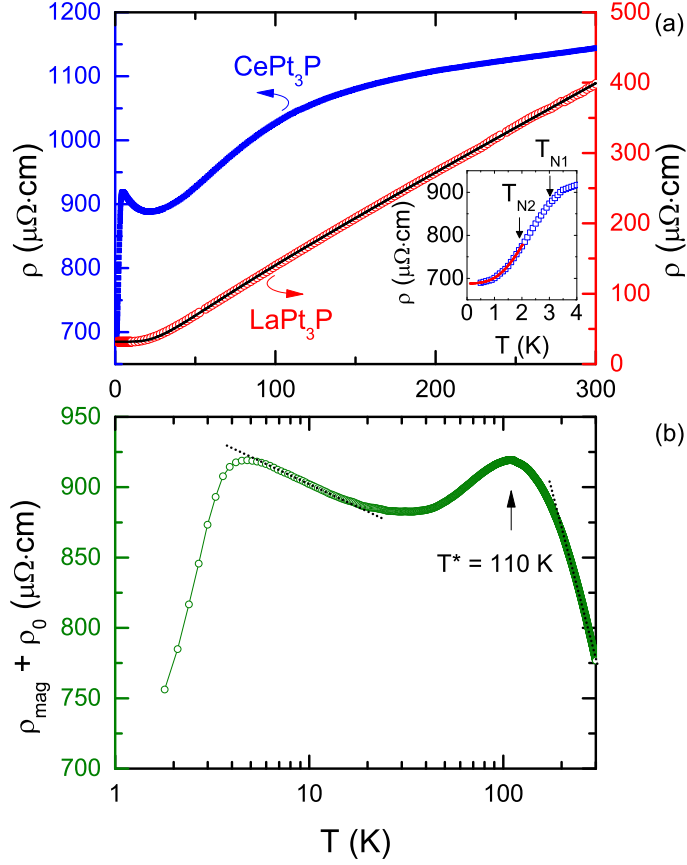}
\caption{\textbf{Transport properties as a function of
temperature.} \textbf{(a)} $\rho(T)$ of CePt$_{3}$P and LaPt$_{3}$P in a
linear temperature scale. The solid line is a fit to the
Bloch-Gr\"{u}neissen-Mott formula (Eq.\ref{eq6}). \textbf{(b)} The magnetic
contribution to the electrical resistivity of CePt$_{3}$P,
$\rho_{mag}$, versus log$T$. The
dashed lines display linear fits in the low and the high
temperature regions, respectively. Inset to (a) plots a fit to
Eq.\ref{eq8} below $T$ $\leq$ 1.9 K. \label{fig5}}
\end{figure}

The temperature variation of the electrical resistivity of
CePt$_{3}$P, $\rho(T)$, is plotted in Figure \ref{fig5}(a). The
resistivity at room temperature is $\rho_{300K}$ = 1140
$\mu\Omega\cdot$cm, a value rather typical for the Ce-based Kondo
compounds with narrow $f$-band \cite{Lohneysen98}. The resistivity
decreases with decreasing temperature and exhibits two features. A
broad hump around 110 K reflects the 4$f$-electron contribution
via Kondo scattering from different CEF levels \cite{CEF1,CEF2}.
At low temperature, a pronounced peak in $\rho(T)$ around 3 K is
directly visible, indicating the AFM ordering phase below $T_{N1}$
= 3.0 K. Above $T_{N1}$, $\rho$ increases in a minus logarithmic
temperature manner over $T$ = 5-20 K, reflecting the Kondo-type
scattering. Further evaluation of $\rho(T)$ requires information
of the phonon contribution which could be taken from the
homologous and isostructural analog, LaPt$_{3}$P. The $\rho(T)$ of
LaPt$_{3}$P, which is also presented in Figure \ref{fig5}(a), can be
well described by a Bloch-Gr\"{u}eneisen-Mott (BGM) relation:
\begin{equation}
\rho(T) = \rho_{0}+4R\Theta_{R}(\frac{T}{\Theta_{R}})^5\int^{\Theta_R/T}_0
\frac{x^5dx}{(e^x-1)(1-e^{-x})}-KT^3,\label{eq6}
\end{equation}
where $\rho_0$ is the residual resistivity due to lattice defects,
the second term denotes electron-phonon scattering, and the third
one accounts for the contribution due to Mott's $s$-$d$ interband
electron scattering. A least square fitting of the BGM formula to
the experimental data over the temperature range 2-300 K leads to
the following parameters: $\rho_0$ = 32 $\mu\Omega\cdot$m,
$\Theta_{R}$ = 160 K, $R$ = 1.25 $\mu\Omega\cdot$cm/K, and $K$ =
4.1$\times$10$^{-8}$ $\mu\Omega\cdot$cm/K$^3$. Note that the
residual resistivity $\rho_0$ is smaller than that in
Ref.\cite{Takayama}. The parameter $\Theta_{R}$ is
usually considered as an approximation of the Debye temperature
$\Theta_{D}$ in spite of some contribution due to electron-electron
correlations in $\Theta_{R}$ \cite{Giovannini}. $\Theta_{D}$
yielded from the specific heat data is 218 K which is in
accordance with $\Theta_{R}$ from the resistivity data. LaPt$_{3}$P
exhibits simple metallic behavior as we expected, without the
characteristic features due to the interplay of Kondo and CEF
effects in CePt$_{3}$P mentioned above.

In order to analyze the magnetic contribution to the electrical
resistivity of CePt$_{3}$P, it is reasonable to assume that the
phonon contribution in this compound can be properly approximated by
that in LaPt$_{3}$P, $\rho_{ph}$ = $\rho(La)$ - $\rho_{0}(La)$, so
we have
\begin{equation}
\rho_{mag}(Ce)+\rho_{0}(Ce)=\rho(Ce)-\rho_{ph}.\label{eq7}
\end{equation}
The temperature dependence of $\rho_{mag}+\rho_{0}$ derived in
this way is presented in Figure \ref{fig5}(b) in a semilogarithmic
scale. As a distinct feature in a Kondo lattice system, a pronounced broad
hump centered at $T^{\ast}$ = 110 K become obvious in $\rho_{mag}$
curve, which could be ascribed to the Kondo scattering from different
CEF levels. According to Cornut
and Coqblin \cite{CEF1}, this maximum provides an estimate of the
CEF splitting energy scale $\sim$ 200 K of Ce-4$f^{1}$ state with
$j$ = 5/2. On the other hand, as temperature is decreased,
$\rho_{mag}$ increases in a logarithmic scale, as shown as the
dotted lines in Figure \ref{fig5}(b) above $T$ $>$ 200 K and between
5-20 K, respectively. Following the theoretical predictions of
Cornut and Coqblin \cite{CEF1}, the logarithmic slopes $c_K^{LT}$
and $c_K^{HT}$ in the low-temperature and high-temperature
regions, respectively, are proportional to the squared effective
degeneracy $\lambda$ of the thermally populated levels: $c_K$
$\propto$ $\lambda^2$-1. For cerium compouns with Ce$^{3+}$ ion
placed in a noncubic crystalline environment the ground multiplet
splits into three doublets, thus the expected ratio is $c_K^{LT}$
: $c_K^{HT}$ = 3:35. In the case of CePt$_{3}$P, with the
coefficients $c_K^{LT}$ = -0.063 and $c_K^{HT}$ = -0.57 yielded
from linear fitting of $\rho_{mag}$ vs log$T$ (see the dashed
lines in Figure \ref{fig5}(b)), the ratio is about 3:27, reasonably
close to the theoretical prediction.

From the inset of Figure \ref{fig5}(a), $\rho$ drops rapidly below
about 3.0 K owing to the reduction of spin-flip scattering upon
entering the AFM ordered state. This magnetic transition
temperature is determined from a slope change of d$\rho$/d$T$ in
Figure \ref{fig2}(b). Upon further cooling, a second slope change in
$\rho$ is observed around 1.9 K, corresponding to the pronounced
kink in d$\rho$/d$T$. Therefore, two magnetic transitions in
CePt$_{3}$P are apparent from the analysis of magnetic
susceptibility $\chi(T)$, specific heat $C(T)$ and electrical
resistivity $\rho(T)$, as shown in Figure \ref{fig2}(b): the first
transition $T_{N1}$ corresponds to the AFM ordering temperature,
while the second one $T_{N2}$ is presumably associated with the
spin reorientation. The values of $T_{N1}$ and $T_{N2}$ derived
from different measurements agree well with each other. It is
noted that while LaPt$_{3}$P shows superconductivity around
$T_{c}$ = 1.0 K (from specific heat), no superconductivity is
observed in CePt$_{3}$P down to 0.5 K.

Considering the relativistic dispersion relation for the AFM
magnon spectrum, the electrical resistivity $\rho(T)$ for $T$
$<$ $\Delta$ can be well described by the following equations
\cite{SDW1,SDW2}:
\begin{align}
\rho(T)=\rho_0+AT^2+B_{\rho}\Delta^2\sqrt{\frac{T}{\Delta}}
[1+\frac{2}{3}(\frac{T}{\Delta})+\frac{2}{15}(\frac{T}{\Delta})^2]\exp(-\Delta/T),
\label{eq8}
\end{align}
where $\rho_0$ is the temperature-independent residual
resistivity, the constant coefficient $B_{\rho}$ is related to the
spin-wave stiffness $D$ by the proportionality $D^{-3/2}$ and
$\Delta$ is the same gap in the spin-wave spectrum as in Eq.(5).
$AT^2$ stems from the electron-electron scattering following the
Fermi liquid theory, while the third term describes the
electron-magnon scattering. This formula is applied to the
electrical resistivity of CePt$_{3}$P (dotted line in the inset of
Figure \ref{fig5}(a)) and a very good fit is obtained with the
fitting parameters: $\rho_0$ = 688 $\mu\Omega\cdot$cm, $\Delta$ =
4.0 K, $A$ = 9.0 $\mu\Omega\cdot$cm/K$^2$ and $B_{\rho}$ = 25
$\mu\Omega\cdot$cm/K$^2$. Considering the relatively short fitting
range of temperature, the derived $\Delta$ value for the
measured polycrystalline sample is still reasonably compared with
that obtained from the specific heat data.

Based on the above analyses, CePt$_{3}$P displays the coexistence
of three important characteristics: AFM ordering of the cerium
local moments due to the  Ruderman-Kittel-Kasuya-Yosida exchange
interaction, the Kondo effect due to the strong $4f$ electron
correlation and moderate effective $4f-5d$ hybridization, and the
CEF interactions. The AFM ordering at $T_{N1}$ = 3.0 K is clearly
identified by the pronounced anomalies in the
temperature-dependent magnetic, thermodynamic and electrical
measurements. In addition, another anomaly at $T_{N2}$ = 1.9 K
is also visible from the physical properties, and is probably
due to a change in the magnetic configuration within the AFM
ordered phase. The behavior of $\rho(T)$ and
$C(T)$ in the ordered region is well describable in terms of AFM
spin-wave spectrum. The field-dependent behavior of the
magnetization and electrical resistivity also indicates a MMT from
the magnetic ordering to a spin-polarized state around
$B_{m}$ = 3.0 T. The magnetic structure of CePt$_{3}$P is still
unclear and the neutron diffraction or M\"{o}ssbauer
spectroscopy experiments are helpful to clarify the details of the
magnetic structure.

The Kondo effect displays itself by the large value of Weiss
temperature $\theta$ (compared with the ordering temperature), the
reduced magnetic entropy and the specific heat jump at $T_N$, as
well as the enhanced Sommerfeld coefficient $\gamma_{Ce}$. From
the analysis of the specific heat data, the Kondo temperature
$T_{K}$ is estimated to be in the range of 2-6 K. Its value can be
also estimated from the magnetic susceptibility as $T_{K}$ $\sim$
$|\theta|/4$ $\simeq$ 7.1 K \cite{TK}, in reasonable agreement with
other estimates. Also, the Kondo effect is well manifested in the
electrical resistivity for Kondo systems with strong CEF
interactions which follows the negative logarithmic-temperature
dependence as $\rho(T)$ = $\rho_0$ + $c_k\ln T$, with Kondo
coefficient $c_k$ $<$ 0 \cite{CEF1}. The inverse susceptibility
(1$/\chi(T)$) curve shows a slope change between $T$ = 100-200 K
which is also attributed to the CEF effect. This temperature
region is in accordance with the energy scale $\Delta_1$ = 240 K
of the multiplet Ce$^{3+}$ ion estimated from the Schottky
contributions of the specific heat \cite{CEF1,CEF2}.

Finally, it is very interesting to compare this CS compound
CePt$_{3}$P with the extensively studied NCS heavy fermion
SC CePt$_{3}$Si ($T_{c}$ = 0.75 K) \cite{CePt3Si}. The crystal
structure of CePt$_{3}$P consists of alternative stacking of
layers of Ce atoms and layers of distorted antiperovskite
Pt$_{6}$P octahedral units along the $c$-axis. The Pt$_{6}$P
octahedra is asymmetrically distorted perpendicular to the
$ab$-plane but alternatively distributed in the $ab$-plane,
resulting in a symmetric antipolar analogue of CePt$_{3}$Si.
CePt$_{3}$Si shows antisymmetric spin-orbit coupling of the
platinum 5$d$ electrons due to the absence of $z$ $\rightarrow$
-$z$ symmetry as well as mixing spin-singlet and spin-triplet
pairing states. The parity mixing alone can hardly account
for the heavy fermion phenomena unless the strong
electron-electron correlation effects which are ensured by
the presence of Ce$^{3+}$ ions are taken into consideration
together \cite{Shiroka}. Correspondingly, the suppression of
superconductivity in CePt$_{3}$P may be attributed to the
enhanced AFM ordering. CePt$_{3}$P is, therefore, probably
placed further away from the magnetic QCP compared with
CePt$_{3}$Si ($T_{N}$ = 2.2 K). With an external control
parameter $\delta$, such as doping or positive pressure, the
system may be shifted towards $T_N$ = 0, namely the QCP
\cite{Stewart,Doniach}. It is thus of great interest to
investigate whether superconductivity exists in  CePt$_{3}$P at
even lower temperature than 0.5 K; if superconductivity does exist,
it will provide strong evidence for the proximity to a magnetic QCP
in CePt$_{3}$P. Comparing with CePt$_{3}$P, the occurrence of
superconductivity at $T_{c}$ = 0.75 K in CePt$_{3}$Si implies that
the NCS crystal structure may favor unconventional superconductivity
within the AFM ground state.

\section*{Conclusion}

In summary, we report the successful synthesis of a new compound
CePt$_{3}$P. From the collected experimental data of magnetization,
specific heat and transport measurements, this compound is
characterized as an antiferromagnetic Kondo lattice with crystal
electric field effect. Two successive magnetic transitions of Ce
4$f$ moments are observed: the magnetic ordering at $T_{N1}$ = 3.0 K
and the spin reorientation at $T_{N2}$ = 1.9 K.
Considering the moderately enhanced Sommerfeld coefficient of
$\gamma_{Ce}$ = 86 mJ/mol$\cdot$K$^{2}$ in the paramagnetic region
and large value of $\gamma_{0}$ = 247 mJ/mol$\cdot$K$^{2}$ in the
the AFM region, the Kondo effect and the AFM order should coexist
in the ground state. Thus a relatively large Fermi surface formed
by the heavy quasiparticles is expected in CePt$_{3}$P with a
Kondo temperature $T_{K}$ $\sim$ 2-6 K. The $ab$ initio
crystal-field and electronic band structure calculations are
necessary to further complement the present results. Further
experiments such as chemical doping are presently underway in
order to tune the ground state from the AFM ordering to
strongly-correlated paramagnetic region.


\section*{Experimental methods}

The polycrystalline sample of CePt$_{3}$P was synthesized by solid
state reaction. Ce piece (99.8$\%$) , Pt powder (99.9$\%$) and P
lump (99.999$\%$) of high purity from Alfa Aesar were used as
starting materials. Firstly, CeP was pre-synthesized by reacting
Ce and P at 1173 K for 72 h. Secondly, powders of CeP and Pt were
weighed according to the stoichiometric ratio, thoroughly ground
and pressed into pellets. The pellets were then packed in
Al$_2$O$_3$ crucibles and sealed in an evacuated quartz tube which
were slowly heated to 1273 K and kept at that temperature for 7
days. Finally, the samples were thoroughly ground, cold pressed
and annealed in vacuum to improve the sample homogeneity. For
comparison, the polycrystalline sample LaPt$_{3}$P was also
synthesized in the similar process. All the preparation procedures
except heating were carried out in an argon protected glove box
with the water and oxygen content below 0.1 ppm. The obtained
CePt$_{3}$P sample is less compact than LaPt$_{3}$P and both of
them are quite stable in the air.

Powder x-ray diffraction (XRD) measurements at room temperature
were carried out on a PANalytical x-ray diffractometer (Model
EMPYREAN) with a monochromatic Cu $K_{\alpha1}$ radiation and a
graphite monochromator. Lattice parameters were derived by
Rietveld refinement using the program RIETAN 2000 \cite{RIETAN}.
The energy dispersion x-ray spectroscopy (EDS) analysis was
performed on a EDS spectrometer affiliated to a field emission
scanning electron microscope (FEI Model SIRION). The electron beam
was focused on a crystalline grain and the chemical compositions
were averaged on at least 4 EDS spectra from different grains. The
electrical resistivity $\rho(T)$ was measured by the standard
four-probe method in a Quantum Design physical property
measurement system (PPMS-9). The
dc magnetization was measured in a Quantum Design magnetic
property measurement system (MPMS-5) with the temperature range of
$T$ = 2-400 K. The specific heat measurements were performed in
the PPMS-9 down to about 0.5 K.

\section*{Acknowledgements}

The authors acknowledge useful discussions with Qimiao Si and
Yongkang Luo. This work was supported by the Ministry of Science and Technology of China (Grants No. 2014CB921203 and 2016YFA0300402) and the National Natural Science Foundation of China (Grants No. U1332209, 11190023 and 11474082).

\section*{Author contributions}

J.C., Z.W., S.Y.Z. and C.M.F. performed the experiment(s), J.C. and Z.W. analyzed the results. J.C and Z.A.X. designed the research and wrote the manuscript. All authors reviewed the manuscript.

\section*{Additional information}

\textbf{Competing financial interests}: The authors declare no competing financial interests.

\newpage

\begin{table}[ht]
\centering
\begin{tabular}{|l|l|l|l|l|}
\hline
  &  SrPt$_{3}$P&  CaPt$_{3}$P&  LaPt$_{3}$P&  CePt$_{3}$P\\
\hline
$a$ (${\rm{\AA}}$) &  5.8094 & 5.6673 & 5.7597 & 5.7123  \\
\hline
$c$ (${\rm{\AA}}$) & 5.2822 & 5.4665 & 5.4736 & 5.4679  \\
\hline
$z_{\rm{Pt(II)}}$ &  0.1409 & - & 0.1459 & 0.1582 \\
\hline
$z_{\rm{P}}$ & 0.7226 & - & 0.7691 & 0.8310 \\
\hline
$T_{c}$/$T_{N}$ (K)& 8.4 & 6.6 & 1.5 & 3.0 ($T_{N1}$)  \\
&&&& 1.9 ($T_{N2}$)\\
\hline
$\rho_{0}$ ($\mu$$\Omega$$\cdot$cm) &  140 & - & 32 & 688  \\
\hline
$\gamma_{0}$ (mJ/mol$\cdot$K$^{2}$)&  12.7 & 17.4 & 1.5 & 86 \\
\hline
\end{tabular}
\caption{\label{arttype} \textbf{Comparisons of physical parameters among
the $A$Pt$_{3}$P family with $A$ = Sr, Ca, La and Ce.} Atomic
positions: $A$ (0, 0, 0), Pt(I) (1/4, 1/4, 1/2), Pt(II) (0, 1/2,
$z_{\rm{Pt(II)}}$), P (0, 1/2, $z_{\rm{P}}$). Note that data
of Sr and Ca are taken from Ref.\cite{Takayama}.}
\label{tab}
\end{table}

\end{document}